\documentstyle[12pt,epsfig]{article}
\begin{document}
\newcommand{\vv}{{\rm v}}
\bibliographystyle{unstr}
\vbox{\vspace{6mm}}
\vspace*{-2cm}
\begin{flushright}
  FIAN-TD22-99 \\
  LPT-9974 \\
  UCRHEP-E264 \\
  October 4, 1999 
\end{flushright}
\medskip
\begin{center}
{\large \bf EVOLUTION OF AVERAGE MULTIPLICITIES OF QUARK AND GLUON JETS \\}
\vspace{2mm}
A. Capella$^1$, I.M. Dremin$^2$, J.W. Gary$^3$,\\ 
V.A. Nechitailo$^2$, J. Tran Thanh Van$^1$ \\
\vspace{2mm}
{\it $^1$LPTHE, Univ. Paris-Sud F-91405 Orsay, Cedex, France \\
$^2$Lebedev Physical Institute, Moscow 117924, Russia    \\
$^3$University of California, Riverside CA 92521, USA } \\
\end{center}
\vspace{2mm}
\begin{abstract}
\noindent
The energy evolution of average multiplicities of quark and gluon jets is
studied in perturbative QCD. Higher order (3NLO) terms in the perturbative
expansion of equations for the generating functions are found.
First and second derivatives of average multiplicities are calculated. 
The mean multiplicity of gluon jets is larger than that of quark jets
and evolves more rapidly with energy.
It is shown which quantities are most
sensitive to higher order perturbative and nonperturbative corrections.
We define the energy regions where the corrections to different quantities
are important. The latest experimental data are discussed.
\end{abstract}

\section{Introduction}
The progress in experimental studies of separated quark 
and gluon jets is very impressive~\cite{bib-reviews}.
Their separation has become possible due
to new methods and high statistics at the $Z^0$ resonance,
which also allows the selection of jets with different sub-energies. 
Therefore the detailed study of the energy evolution of such properties
of jets as their average multiplicities and widths has also become possible. 
It is well known
that the average multiplicities of quark and gluon jets increase quite rapidly
with energy but their ratio has a much slower dependence and 
tends to a constant in asymptotics. 
The energy dependence of this ratio and of the ratio of the
derivatives of mean multiplicities is of much debate nowadays,
with new experimental data appearing rather regularly. 
The experimental procedure of the jet separation, the scale choice,
and the manner in which soft particles are ascribed to jets,
is often disputable, however.
Therefore,
in the following,
we do not insist upon a direct quantitative comparison of data
with theory but rather describe the correspondence between 
our present experimental and theoretical findings
with the purpose of finding
quantities sensitive to areas of disagreement.

On the theoretical side, perturbative QCD (pQCD)
provides quite definite
predictions which can be confronted to experiment. 
The asymptotical value of
the ratio of average multiplicities in gluon and quark jets 
equals 2.25~\cite{brod},
much larger than its
experimental values which range from about 1.04 at the
comparatively low energies of the $\Upsilon$ resonance~\cite{cleo97}
to about 1.5 at the $Z^0$ resonance~\cite{opal96,opal99}.
The next-to-leading order corrections to this ratio $r$ 
(we call them NLOr) reduce it from its asymptotical value 
by about 10$\%$ at the $Z^0$ energy~\cite{mu, mw, cdfw}. 
The next order NNLOr
terms diminish it further,
although calculations by different 
methods~\cite{gaff, dr3} lead to different results as we discuss later. 
We have calculated the next-to-next-to-next-to-leading order 
(3NLOr) terms using QCD equations for the generating functions. 
The 3NLOr terms are small but nonetheless
provide an improved description of data,
as shown below.

The computer solution of QCD equations for the generating functions has shown
even better agreement with experiment for this ratio~\cite{lo},
and excellent agreement for higher moments of the 
multiplicity distributions as well~\cite{lup}.
Being perfect at the $Z^0$ energy,
the agreement for the ratio is less good at the $\Upsilon$ resonance
where the theoretical curve is 15-20$\%$ 
above the experimental result (see Fig.~2 in~\cite{lo}). 
In other words, the theoretically predicted
{\em slope of the ratio} of multiplicities in gluon and quark jets is
noticeably smaller than its experimental value. 
Nonetheless,
there has been a steady convergence of theory and experiment 
as more accurate calculations and more sophisticated measurements
have become available.
Moreover, it is even surprising that any agreement
is achieved given that the expansion parameter is extremely large
(about~0.5) at present energies.

Recently, experimental data about {\em the ratio of slopes} of average
multiplicities in gluon and quark jets have been reported~\cite{lan}
and the corresponding analytical expressions have been 
calculated~\cite{dr1}.
The importance of studying the slopes stems from the fact that 
some of them are extremely sensitive to higher order 
perturbative corrections,
and to nonperturbative terms at available energies,
as is discussed below.
Thus they provide an opportunity to learn more
about the structure of the perturbative series from experiment. 

The experimental group~\cite{lan} claims
that it is possible to simultaneously describe the scale
dependence of multiplicities in gluon and quark jets
only if nonperturbative
effects contribute constant terms to the multiplicities.
We examine this issue.
The success of the computer solution
suggests an alternative possibility,
namely that a combination of nonperturbative terms which rapidly 
decrease with increasing energy and higher order perturbative terms
can describe the data,
effectively replacing the constant offsets assumed in~\cite{lan}.
To verify this, 
one needs analytical calculations of the corrections for
experimental quantities most sensitive to them. 
With this aim in mind, 
we consider analytical expressions for the above ratios, 
as well as for the ratio of the second derivatives (curvatures)
of average multiplicities.
These last ratios have not yet been measured. 
The general problem of perturbative and nonperturbative
terms in the QCD equations for the generating functions is also
addressed using simple analytical estimates. 
These estimates show that, at asymptotically high energies,
the role of the nonperturbative region is not as important as the higher
order perturbative terms but that it
becomes more important at lower energies.
The location of the borderline depends on the quantity under consideration.

For the comparison of theory with data,
we utilize the principle of 
local parton hadron duality (LPHD)~\cite{bib-lphd},
i.e.~the hypothesis that a hadron distribution (experiment)
can be related to a parton distribution (theory)
by a simple normalization constant.
LPHD has been shown to be reliable for distributions dominated
by soft particles,
such as the multiplicity related quantities studied here
(for a fuller discussion, see~\cite{bib-ochs}).

\section{Evolution of mean multiplicities in pQCD}
\label{sec-means}

In perturbative QCD, 
the general approach to studying multiplicity distributions 
is formulated in the framework of equations for generating
functions (for reviews see~\cite{dkmt,dr2}). 
From the generating functions, 
one derives equations for average multiplicities and, 
in general, for any moment of the multiplicity
distributions~\cite{dr3,d1,dhwa}. 
In particular, the equations for the average
multiplicities of gluon and quark jets are
\begin{eqnarray}
\langle n_G(y)\rangle ^{'} =\int dx\gamma _{0}^{2}[K_{G}^{G}(x)
(\langle n_G(y+\ln x)\rangle +\langle n_G(y+\ln (1-x))\rangle -\langle n_G(y)
\rangle ) \nonumber  \\
+n_{f}K_{G}^{F}(x)(\langle n_F(y+\ln x)\rangle +\langle n_F(y+
\ln (1-x))\rangle -\langle n_G(y)\rangle )],  \label{ng}
\end{eqnarray}
\begin{equation}
\langle n_F(y)\rangle ^{'} =\int dx\gamma _{0}^{2}K_{F}^{G}(x)
(\langle n_G(y+\ln x)\rangle +\langle n_F(y+\ln (1-x))\rangle -\langle n_F(y)
\rangle ).   \label{nq}
\end{equation}
From these equations one can predict the energy evolution
of the ratio of multiplicities $r$ and the QCD anomalous dimension $\gamma$
(the slope of the logarithm of the
average multiplicity in a gluon jet),
defined as
\begin{equation}
r=\frac {\langle n_G\rangle }{\langle n_F\rangle }\; ,\;\;\;\;\; \;\;\;
\gamma =\frac {\langle n_G\rangle ^{'}}{\langle n_G\rangle }
=(\ln \langle n_G\rangle )^{'}\; .  \label{def}
\end{equation}
Here, the prime denotes the derivative 
with respect to the evolution parameter
$y=\ln (p\,\Theta /Q_{0})$,
$p$ and $\,\Theta $ are the momentum and initial angular
spread of the jet related to the parton virtuality 
$Q=p\,\Theta /2$ (for small angles),
\, $Q_{0}$=const, \, $K$'s are the well known splitting
functions, $\langle n_G\rangle $ and
$\langle n_F\rangle $ are the average multiplicities in gluon and quark jets, \,
$\langle n_G\rangle ^{'}$ is the slope of $\langle n_G\rangle $, 
and $n_f$ is the number of active flavors. 
The perturbative expansions of $\gamma $ and $r$ are
\begin{equation}
\gamma =\gamma _{0}(1-a_{1}\gamma _0 -a_{2}\gamma _{0}^{2} -a_{3}\gamma _{0}^{3}
)+O(\gamma _{0}^{5}),  \label{gam}
\end{equation}
\begin{equation}
r=r_0(1-r_{1}\gamma _{0}-r_{2}\gamma _{0}^{2}-r_{3}\gamma _{0}^{3})+O(\gamma
_{0}^{4}),   \label{rat}
\end{equation}
where $\gamma _{0}=\sqrt {2N_{c}\alpha _{S}/\pi }, \, \alpha _{S}$
is the strong coupling constant,
\begin{equation}
\alpha_{S}=\frac {2\pi }{\beta _{0}y}\left [1-\frac {\beta _{1}\ln (2y)}
{\beta _{0}^{2}y}\right ]+O(y^{-3}),   \label{alp}
\end{equation}
$\beta _{0}=(11N_{c}-2n_{f})/3, \, \beta _{1}=[17N_{c}^2- n_{f}(5N_c+3C_F)]/3,
r_0 = N_c/C_F,$ \,and in QCD $N_{c}=3$ is the number of colors
and $C_{F}=4/3$.

The limits of integration in eqs.~(\ref{ng}), (\ref{nq}) are generally
chosen either to be 0 and 1 or $e^{-y}$ and $1-e^{-y}$.
This difference, being negligible at high energies $y$, 
becomes more important at low energies. 
Moreover, it is of physical significance. With limits
of $e^{-y}$ and $1-e^{-y}$, the partonic cascade terminates at
virtuality $Q_0$ as seen from the arguments of the multiplicities in the
integrals. With limits of 0 and 1, the cascade extends into the
nonperturbative region with virtualities smaller than $Q_{0}/2$, 
i.e.~$Q_{1}\approx xp\Theta /2$ and $Q_{2}\approx (1-x)p\Theta /2$,
where the very notion of partons becomes scarcely applicable. 
This region contributes
terms of the order of $e^{-y}$, power-suppressed in energy. 
It is not clear whether the equations and the LPHD hypothesis
are valid down to the cutoff $Q_0$ only
or if the nonperturbative region can be included as well.

Nonetheless, the purely perturbative expansions (\ref{gam}) and (\ref{rat})
with constant coefficients $a_{i}, r_{i}$ and energy-dependent $\gamma_0$
are applicable only in the case of limits 0 and 1. This is quite natural for
infrared-safe quantities like mean multiplicities. According to (\ref{def}),
each derivative gives rise to the factor $\gamma$. Therefore, a Taylor
series expansion of eqs.~(\ref{ng}) and (\ref{nq}) is equivalent~\cite{d1}
to a perturbative series. Terms of the same order should be equal on
both sides. From this, one obtains the coefficients $a_i, r_i$.
The analytical formulas for
$a_i , r_i$ with $i\leq 3$ are given in Appendix~1. 
For $i\leq 2$, they were presented earlier~\cite{mu,mw,dr3}.
We present them here using a more general notation.
The numerical values of $a_i , r_i$ for different numbers of 
active flavors $n_f$ and in SUSY QCD~\cite{dkmt}
are given in Table~\ref{tab-randa}.

\begin{table}[h]
\begin{center}
\begin{tabular}{|c|c|c|c|c|c|c|}
\hline
$n_f$ & $r_1$ & $r_2$ & $r_3$ & $a_1$ & $a_2$ & $a_3$\\
\hline
3 &  0.185 & 0.426 & 0.189   & 0.280 & - 0.379 & 0.209\\
\hline
4  &  0.191 & 0.468 & 0.080  & 0.297 & - 0.339 & 0.162 \\
\hline
5 &    0.198 & 0.510 &  -0.041  & 0.314 & - 0.301 & 0.112\\
\hline
S &  0     &  0    &    0    & 0.188 & -0.190  & -0.130  \\
\hline
\end{tabular}
\end{center}
\caption{Numerical values
of the perturbative corrections up to order $\gamma _{0}^{3}$
for the multiplicity ratio $r$ and the QCD anomalous dimension~$\gamma$,
based on integration limits of the generating functions
from 0 to~1 (see text);
$n_f$ is the number of active quark flavors while S refers
to SUSY QCD.
}
\label{tab-randa}
\end{table}
At the $Z^0$ resonance, 
for which the energy of an unbiased gluon jet 
is about 40~GeV~\cite{opal96,opal99},
the subsequent terms diminish the value of $r$ compared 
with its asymptotic value $r_0 =2.25$ by
approximately 10\%, 13\%, and 1\% for $n_f=4$. 
The smallness of $r_3$ is
a good indication of the convergence of the series at this energy.

Note that we have used the traditional method (see~\cite{dkmt}) 
of considering $r$ and $\gamma$ to be the main ingredients
of the calculation and that we thus employ the conventional
definition for the order of the corrections.
This method introduces some asymmetry
between quark and gluon jets. 
A more natural way to structure the calculation
might be to define the anomalous dimensions 
of gluon and quark jets symmetrically, as
\begin{equation}
  \gamma =\frac {\langle n_G\rangle '}{\langle n_G\rangle }=[\ln \langle n_G
  \rangle ]' ; \;\;\; \gamma _F= \frac {\langle n_F\rangle '}{\langle n_F\rangle }=
  [\ln \langle n_F\rangle ]'  \label{ggf}
\end{equation}
and to call the subsequent corrections next-to-leading-order (NLO), etc.
With the conventional approach used here, 
it is necessary to differentiate between corrections to $r$
(which we call NLOr, etc.)
and those to~$\gamma$.
For example, the NLOr-correction
given by $r_1$ contributes to $\gamma _F$
only at NNLO and higher orders
(for more details see the next section). 
The differences between the traditional and symmetric
methods are not important, however, as long as
the corrections themselves are small.

The behavior of the ratio $r$ in various approximations 
(DLA=leading order, NLOr, NNLOr, 3NLOr) 
is shown in Fig.~1 for $n_f=4$.
The abscissa variable is $\kappa =p\Theta \approx 2E\sin (\Theta/2)$,
where $y \approx \ln (\kappa /Q_0)$.
Here $Q_0$ is chosen to depend on the number of active flavors in 
the proportions used previously~\cite{dhwa} 
in order that $\alpha _S$ agree with experiment. 
A comparison of these results to data is given
in section~\ref{sec-experiment}.

Now we would like to return to the issue of the limits of integration
mentioned above. One can analytically estimate how the coefficients $a_i, r_i$
are modified if the range of integration becomes somewhat narrower, e.g.
from $\varepsilon $ to $1-\varepsilon $ ($\varepsilon =$const $\ll 1$).
In that case one easily calculates the corrections to $a_i, r_i$.
These are given 
in the integrals $\vv_i$ presented in Appendix~1. In particular,
using the formulas of Appendix~1 for $\varepsilon =0.1,\; n_f=4$,
one obtains
rather large contributions from the nonperturbative region:
$a_{1}^{NP}=-0.36a_1, \; r_{1}^{NP}=-0.56r_1$. 
However $e^{-y}=0.1$ only at the $\Upsilon $-resonance and
is much smaller at the $Z^0$. 
Therefore, narrowing the range of integration reduces
the energy dependence.
The 3NLO result for $r$ based on the narrower range of integration
is shown by the curve labeled $r(\varepsilon)$ in Fig.~1.
Surely, it is not entirely
satisfactory to replace $e^{-y}$ in the limits of integration by a constant
$\varepsilon $ because if $\varepsilon $ depends
on $y$ the coefficients $a_i, r_i$ become energy dependent and their derivatives
need to be considered as well. 
Nonetheless such a procedure can be used
to obtain an order-of-magnitude estimate. 
To obtain better accuracy in the energy region
from the $\Upsilon $ to the $Z^0$,
the computer solution of eqs. (1), (2) with modified (in the nonperturbative
region) coupling strength, higher order terms in the kernels
and more strict conservation laws (in particular, of transverse momenta)
is necessary.

\section{Slopes as a stumbling-block of pQCD}
\label{sec-slopes}

We have shown how the perturbative series can be used to evaluate
$\gamma$ and $r$, and where nonperturbative
corrections become important. 
In this section, we show that the perturbative
approach should work well for the ratio of the slopes of the
multiplicities but that it is much less reliable to use low order
perturbative estimates,
even at the $Z^0$-energy, 
for such quantities as the slope of $r$ 
or the ratio of slopes of logarithms of
multiplicities (we shall call them the logarithmic slopes). 
Much larger energies are needed for that. 
Thus the values of $r^{'}$ and/or of
the logarithmic slopes can be used
to verify the structure of the perturbative expansion.

The slope $r'$ is extremely sensitive to higher
order perturbative corrections. The role of higher order corrections is
increased here compared to $r$ because each $n$th order term proportional
to $\gamma _{0}^{n}$ gets an additional factor $n$ in front of it when
differentiated, the main constant term disappears, 
and the large value of the ratio $r_2/r_1$ becomes crucial:
\begin{equation}
r^{'} =Br_{0}r_{1}\gamma _{0}^{3}\left [1+\frac {2r_{2}\gamma _{0}}
{r_1}+\left (\frac {3r_3}{r_1}+B_{1}\right )\gamma _{0}^{2}+O(\gamma _{0}^{3})
\right ],       \label{rpri}
\end{equation}
where  the relation $\gamma _{0}^{'}\approx -B\gamma _{0}^{3}(1+B_{1}\gamma
_{0}^{2})$ has been used with $B=\beta _{0}/8N_c =1/2C^2 ;
B_{1}=\beta _{1}/4N_c\beta _0$. The factor in front of the bracket is 
very small even
at present energies: $Br_0r_1\approx 0.156$ and $\gamma _0\approx 0.5$.
Nonetheless, the numerical estimate of 
$r^{'}$ is unreliable because of the expression in the brackets.
Each differentiation leads to a factor $\alpha _S$ (or $\gamma _{0}^{2}$), 
i.e.~to terms of higher order.
For the values of $r_1$, $r_2$, $r_3$ given in Table~1 ($n_f=4$),
one estimates
$2r_{2}/r_{1}\approx 4.9$, $(3r_3/r_1)+B_1\approx 1.5$ (see eq. (\ref{rnum})
and note the aforementioned factors of 2 and 3 in front of $r_2$ and $r_3$,
correspondingly).
The simplest correction, proportional to $\gamma _0$,
is more than twice as large as 1
at present energies,
while the next correction,
proportional to $\gamma _0^2$,
is about~0.4. 
Therefore even higher order terms should be calculated 
before the perturbative result for $r'$ can be considered
to be reliable.

The slope $r'$ equals 0 for a fixed coupling constant.
In the case of the running coupling constant, 
it evolves rapidly with $y$ in
the lowest perturbative approximations (\ref{rpri}) 
(this is also seen from the Figs. of \cite{lo}), 
and ranges according to (\ref{rpri}) from 0.06
at the $Z^0$ to 0.25 at the $\Upsilon $. 

Let us consider the ratios of slopes $r^{(1)}$ and curvatures $r^{(2)}$
defined as
\begin{equation}
r^{(1)}=\frac {\langle n_G\rangle ^{'}}{\langle n_F\rangle ^{'}}; \,\,\,
r^{(2)}=\frac {\langle n_G\rangle ^{''}}{\langle n_F\rangle ^{''}}. \label{r12}
\end{equation}
Their ratios to $r$ can be written as
\begin{equation}
\rho _1 =\frac {r}{r^{(1)}}=1-\frac {r^{'}}{\gamma r},   \label{rh1}
\end{equation}
\begin{equation}
\rho _2 =\frac {r}{r^{(2)}}=1-\frac {2\gamma rr^{'}+rr^{''}-2r^{'2}}
{(\gamma ^{2}+\gamma ^{'})r^2}.   \label{rh2}
\end{equation}
Since $r^{'} \propto \gamma _{0}^{3}$, the asymptotical values of $r,\,r^{(1)},\,
r^{(2)}$ coincide and should equal~2.25. Moreover, the values of
$r,\; r^{(1)},\; r^{(2)}$ coincide in NLOr. 
This implies that the mean multiplicities
in gluon and quark jets and their derivatives differ in NLO by a constant
factor $r_0$ only.
First preasymptotical corrections are very small.
They are of the order of $\gamma _{0}^{2}$ with a small
factor in front and contribute about 2--4$\%$ at  present energies:
\begin{equation}
\rho _1 =1-Br_{1}\gamma _{0}^{2}\approx 1-0.07\gamma _{0}^{2},  \label{cor1}
\end{equation}
\begin{equation}
\rho _2 =1-2Br_{1}\gamma _{0}^{2}\approx 1-0.14\gamma _{0}^{2}.
\label{cor2}
\end{equation}
However, this ideal perturbative situation 
does not persist when the next terms are
calculated (see Appendix~2). For example, for $\rho _1$ their numerical
values are
\begin{equation}
\rho _1 =1-0.07\gamma _{0}^{2}(1+5.38\gamma _{0}+4.14\gamma _{0}^{2}) \;\;
(n_f =4).   \label{cor3}
\end{equation}
All the correction terms are much smaller than~1.
However,
the numerical factors inside the brackets are so large that at present values
of $\gamma _{0}\approx 0.5$ the subsequent terms are larger than the first one
and the sum of the series is unknown. One can sum the series
using the simplest Pade-approximant
(i.e. by assuming the steady decrease of the unaccounted terms 
in proportions determined by the ratio of two last calculated terms), 
resulting in
\begin{equation}
\rho_1=1-0.07\gamma_0^2\left [1+\frac {5.38\gamma_0}{1-0.78\gamma_0}\right ].
\label{rpad}
\end{equation}

In contrast to $\rho_1$,
the perturbative corrections to
the ratio of slopes $r^{(1)}$ are small.
The lowest order $O(\gamma_0)$ correction
to $r^{(1)}$ is the same as for $r$ but higher order
corrections are smaller because they are negative both in $r$ and $\rho_1$
which define $r^{(1)}=r/\rho_1$. 
This explains why experimental values of $r^{(1)}$
are similar to values of $r$ calculated in the NLOr-approximation 
(see section~\ref{sec-experiment}),
whereas experimental values
of $r$ are still about 25$\%$ lower than the NLOr-prediction.
Yet more substantial cancelations of higher order terms 
should occur for~$r^{(2)}$.
It would be interesting to check this prediction experimentally.

The value of $\rho _1$ also determines the ratio of the anomalous dimensions,
i.e. of the  slope of the logarithm of the average
multiplicity of a quark jet ($\gamma _F$) to that of a gluon jet ($\gamma $):
\begin{equation}
\gamma_F =\rho_1 \gamma =\gamma -\frac {r^{'}}{r}.
\end{equation}
The logarithmic slopes of quark and gluon jets are equal in NLO.
They differ in higher orders in such a manner that 
$\gamma _F<\gamma $ since both $r$ and $r^{'}$ are positive.
Our failure to obtain a precise estimate of $\rho _1$ implies that
we can not perturbatively evaluate $\gamma _F$ either.
The ratio of logarithmic slopes $\rho_1$
is more sensitive to higher order perturbative terms
than the ratio of slopes $r^{(1)}$,
similar to~$r^{'}$.

An interesting feature of eq.~(\ref{cor3}) is that $\rho _1$ contains
terms up to $O(\gamma _0^4)$ whereas the $r$ in the 
numerator of (\ref{rh1}) is known to $O(\gamma _0^3)$ only. 
This situation is related
to the fact that $r'/\gamma \sim O(\gamma _0^2)$ and
allows the perturbative expression for $\gamma _F$ to be written as:
\begin{eqnarray}
 \gamma _F & = & \gamma _0 \left[ \right. 1-a_1\gamma _0-(a_2+Br_1)\gamma _0^2
            -(a_3+2Br_2+Br_1^2)\gamma _0^3   \nonumber \\ 
  & & -( a_4+B(3r_3+3r_2r_1+B_1r_1+r_1^3))\gamma _0^4 \left. \right]  
   \label{gf4}
\end{eqnarray}
if the higher order 4NLO term $-a_4\gamma _0^4$ is added to $\gamma $ inside
the brackets in (\ref{gam}). The value of $a_4$ is not yet known. The formula
(\ref{gf4}) clearly displays the relation between 
the NLO and NLOr approximations.
One sees that the NLOr term $r_1$ is summed
with the NNLO value of $a_2$. 
The analogous situation occurs for higher order terms. 
In particular, the 4NLO term contains, beside $a_4$,
the contributions from the 3NLOr, NNLOr and NLOr approximations.

We conclude that within the present accuracy of $O(\gamma_0^3)$ corrections,
the perturbative QCD approach fails in the precise determination of the
logarithmic slope of the quark jet multiplicity $\gamma _F$
even at the $Z^0$ and can be trusted only at much higher energies.

\section{Theory and experiment on multiplicities and slopes}
\label{sec-experiment}

Experimental results on multiplicities and slopes
in quark and gluon jets are available in the energy range 
from the $\Upsilon$ to the $Z^0$ \cite{cleo97,opal99,lan}. 
The CLEO data at the $\Upsilon$~\cite{cleo97} and
the OPAL data at the $Z^0$~\cite{opal99}  
are fully inclusive (unbiased jets).
Thus, the jet definitions correspond closely to theory.
The DELPHI data~\cite{lan} 
at intermediate scales to OPAL and CLEO are based 
on three jet events selected using a jet finder.
Parameters of the DELPHI distributions are given in
terms of the jet hardness scale $\kappa =2E\sin (\Theta /2)$, 
where $\Theta $ is the opening angle.
For small $\Theta $, one gets $y \approx \ln (\kappa /Q_0)$
as mentioned above. 
The data of CLEO and OPAL are presented in terms of
the jet energy or invariant mass.
Leaving aside the question
of whether the DELPHI results can be directly compared with pQCD
and the difference in scales at large $\Theta $, we proceed
to an analysis keeping in mind that qualitative trends and 10--20$\%$
accuracy will satisfy us at present.

The experimental results for $r$ are shown in Fig.~1.
The data are seen to lie substantially below the predictions.
The theoretical results approach the experimental values 
more closely, however,
as higher order terms are included.
The comparatively large value of $r_2$ helps in this regard
although it also poses some problems for the perturbative
treatment of slopes,
as discussed in section~\ref{sec-slopes}.
Agreement within 10-20\% accuracy is achieved 
for the OPAL value of $r$ at the~$Z^0$. 
Here $E\approx 40$~GeV and $\kappa \approx 37$~GeV 
is only slightly smaller than $E$.
With $\alpha_S (37$ GeV$)\approx 0.14$ and 
$\gamma_0 \approx 0.517$ one obtains 
from (\ref{rat}) $r \approx 1.71$ for $n_f=3$ and $r \approx 1.68$ for $n_f=4$,
i.e. much smaller values than 2.25 at asymptotics.  The computer
solution \cite{lo} leads to perfect agreement with the OPAL result $r_{exp}
\approx 1.51$.
The computer solution differs from 
analytical estimates by use of the explicit multiplicities in the integrals,
with no Taylor series expansion. 
This fact becomes especially
important for higher order terms at present values of the expansion parameter
$\gamma _0$. The limits  $e^{-y}$ and
$1-e^{-y}$ of integration and the scale of the running coupling constant
(however, in the one-loop approximation only) are included.

According to eq.~(\ref{rat}),
the ratio $r$ becomes smaller with decreasing scale
since $\gamma _0$ increases. 
The same is observed for the experimental results. 
Thus the qualitative trends agree. 
The computer solution~\cite{lo} decreases more
slowly than the experimental data,
exceeding the experimental value of $r$ 
by 15--20$\%$ at the $\Upsilon $
as mentioned in the introduction.
This results in a noticeable difference between the
theoretical (computer) and experimental values of $r^{'}$. 
From Fig.~2 in~\cite{lo}, 
the slope predicted by the computer solution 
can be estimated to be about~0.096
at the scale of the Z$^0$ .
From the data in Fig.~1,
the corresponding result is about~0.174.
These values differ from the analytic prediction
of~0.06 given in section~\ref{sec-slopes}. 
This indicates the strong influence
of higher order perturbative corrections,
even at the $Z^0$,
and of nonperturbative terms at lower energies which
reduce the energy dependence.

We have also performed a fit of the gluon jet multiplicity.
The only dependence on energy in (\ref{gam}), (\ref{rat})
comes from $\alpha _S(y)$ which we constrain 
to agree with the PDG results~\cite{pdg}. 
The energy dependence of the gluon jet
multiplicity at 3NLO is given by~\cite{dg}
\begin{equation}
\langle n_G\rangle =Ky^{-a_1C^2}\exp 
   \left[ 2C\sqrt y +\delta _G(y) \right], \label{ngy}
\end{equation}
with $K$ an overall normalization constant, $C=\sqrt {4N_c/\beta _0}$, and
\begin{equation}
\delta _G(y)=\frac {C}{\sqrt y}\left [ 2a_2C^2+\frac {\beta _1}{\beta _0^2}
[\ln (2y)+2]\right ] 
+\frac {C^2}{y}\left [ a_3C^2-\frac {a_1\beta _1}{\beta _0^2}
[\ln (2y)+1]\right ].   \label{del}
\end{equation}
The normalization of the gluon jet multiplicity is arbitrary. 
It is determined by the lower limit of integration $y_0$ in
\begin{equation}
  \langle n_G\rangle \propto \exp \left[ \int ^y_{y_0}\gamma (y')dy'\right],  
  \label{nin}
\end{equation}
which is not defined.
We perform a one parameter fit of eq.~(\ref{ngy}) to the gluon
jet data in~\cite{lan},
with $K$ the fitted parameter.
Generally good agreement with this data
is found as shown in Fig.~2.
The rate of the multiplicity increase is somewhat
overestimated by the theory, however,
especially by the 3NLO (dashed) curve. 
The power-like terms provided by the nonperturbative region 
(i.e. the truncation of the
integration in $\vv_i$ at $e^{-y}$ and at $1-e^{-y}$)
are not very important.
They slightly flatten the energy dependence shown in Fig.~2.

One can write the energy dependence of the quark jet multiplicity 
analogously to (\ref{ngy}) as~\cite{dg}
\begin{equation}
  \langle n_F\rangle =\frac {K}{r_0}y^{-a_1C^2}\exp 
     \left[ 2C\sqrt y +\delta _F(y)\right], 
\label{nfy}
\end{equation}
with
\begin{equation}
\delta _F(y)=\delta _G(y)+\frac {C}{\sqrt y}r_1+\frac 
  {C^2}{y}(r_2+\frac {r_1^2}
  {2}).   \label{dfy}
\end{equation}
If the normalization factor $K$ is determined by the fit to
the gluon jet data as in Fig.~2,
the normalization of the quark jet multiplicity
according to formulas (\ref{def}), (\ref{gam}), (\ref{rat})
is too low by about 2--3 units,
as shown in Fig.~2.
This is due to our failure to describe the ratio $r$
with sufficient accuracy as discussed above.
More important is the fact that were we to normalize
$\langle n_F\rangle$ by brute force at some energy, 
we would be unable to describe its slope.
This can be seen from Fig.~2.
The quark jet curves with normalization fixed by the
gluon jet data are seen to have about the 
same slope as the data.
Thus,
if the quark jet predictions are multiplied by a
constant factor so that the experimental and theoretical
results for $\langle n_F\rangle$ agree at some scale~$\kappa$,
the experimental and theoretical slopes would
differ drastically.
In particular,
the solid curve at NLO (i.e. for $\delta _F=0$) 
would be much steeper than the experimental data
and the 3NLO curve would be steeper yet. 
Nonperturbative corrections would flatten the slopes and
yield better agreement between data and theory.
Nonetheless the slope of $\langle n_F\rangle $ is less well
determined than that of $\langle n_G\rangle $.
It is more sensitive to higher order corrections 
(see section~\ref{sec-slopes}).

We should emphasize that there are ambiguities with the
data in Fig.~2.
These concern the dependence on
a jet finding algorithm
and the choice of the scale variable $\kappa$.
Qualitatively similar results to those shown in Fig.~2 are obtained
using unbiased gluon and quark jets which do not
share these ambiguities~\cite{dg}.
For the unbiased jets,
the growth of multiplicity with scale is more marked than
that exhibited by the data in Fig.~2 (see Figs.~1 and~2 in~\cite{dg}).
The 3NLO prediction~(\ref{ngy}) is found to describe the unbiased
gluon jet measurements well,
including the slope,
with a value of $\alpha_S$ consistent with measurement~\cite{pdg}.
Thus the disagreement between experiment
and theory seen in Fig.~2 for the gluon jet slope
is possibly due to the lack of correspondence between the
theoretical and experimental definitions of the jets.
Concerning quark jets,
it was shown in~\cite{dg} that eq.~(\ref{nfy})
describes the unbiased quark jet measurements well
using a value of $Q_0$ about three times smaller
than that found for unbiased gluon jets.
Such a scale is physically acceptable and leads to a
reasonable value of~$\alpha_S$ .
In contrast,
a fit of the quark jet data in Fig.~2
using the technique described in~\cite{dg}
yields $Q_0\sim 10^{-6}$~GeV,
an unphysical value of the scale.

The ratio of slopes $r^{(1)}$ is less sensitive to
higher order corrections and is better 
approximated by the NLOr expression than $r$,
as discussed in section~\ref{sec-slopes}.
According to DELPHI~\cite{lan}, the experimental value $r^{(1)}_{exp}$ 
depends very little on scale and stays almost constant
at about 1.97$\pm $0.10, whereas $r^{(1)}_{NLO}\approx
2.01-2.03$. Using formulas (\ref{rat}), (\ref{rh1}), (\ref{cor3}) one gets
$r^{(1)}_{3NLO}\approx 1.86-1.92$ from the $\Upsilon $ to the $Z^0$. 
Fig.~3 displays
the values of $r^{(1)}$ in the NLOr and 3NLOr-approximations for $n_f=4$
in comparison to the experimental limits~\cite{lan}. 

The measured values of $\rho_{1,exp} \approx r_{exp}/2$
range from 0.53 at the $\Upsilon $ to 0.75 at the $Z^0$. 
Its perturbative values according to
(\ref{cor3}) range from 0.85 to~0.9. Either next order perturbative
or nonperturbative terms are needed \cite{dr1} to reproduce the big difference
$d_1 =r^{(1)}_{exp}-r_{exp}\approx 2(1-\rho_{1,exp})$ observed in experiment.
The simplest Pade expression (\ref{rpad}) improves the situation only slightly.

There is a widely held opinion~\cite{web,sch}
that the average multiplicity in $e^{+}e^{-}$
collisions is well described by pQCD in the NLO-approximation. 
At the same time,
one does not usually mention that the NLO-approximation
fails badly in its prediction for $r$. However 
these two facts are tightly connected if one analyzes 
multiplicities in separated quark 
and gluon jets. As the simplest approach, one used to consider the formula 
$n_{e^{+}e^{-}}(2E)=2n_F(E)$ and, more importantly, to fit its energy behavior
using the anomalous dimension $\gamma$ 
whereas $\gamma_F=\rho_1 \gamma $ should be used. 
In NLO, their ratio $\rho_1$ equals 1 so that this replacement
seems justified were it not for 
the completely wrong (compared with experiment) 
NLO-value of $r$ in the definition of $\rho_1=r/r^{(1)}$,
whereas $r^{(1)}$ in the NLO-approximation approximately 
equals its experimental value $r^{(1)}_{exp}$. 
Therefore, one should 
view this success with some scepticism. 
This issue is also discussed in~\cite{dg}.

This problem revealed itself in experiment when 
the DELPHI data on the scale dependence of
multiplicities in quark and gluon jets became available~\cite{lan}. 
The quark jet logarithmic slope $\gamma _F$ was found 
to be much smaller than the gluon slope $\gamma$, 
and their ratio $\rho_1$ was measured to be
about 0.50--0.75 as mentioned above.
If one insists on using NLO expressions,
it is therefore necessary to change $\langle n_F\rangle $ 
while leaving $\langle n_F \rangle ^{'}$ unchanged. 
This was done in~\cite{lan} by adding an ad-hoc
constant to $\langle n_F\rangle $
which was found to be rather large (about 2--3).
 
It seems premature to introduce constant offsets to the
quark and gluon jet multiplicities before other,
better motivated possibilities are studied both from the
theoretical and experimental sides.
Apart from the ambiguities in the DELPHI measurements discussed above,
there is a rather regular way to account for higher order
perturbative and nonperturbative corrections 
which leads in the proper direction. 
This is supported by the computer solution~\cite{lo} of 
eqs.~(1), (2) with quite satisfactory agreement with 
experiment for the value of $r$ at the $Z^0$. 
The slight difference at low energies results in a disagreement on
slopes which becomes a stumbling-stone for the whole approach 
and requires very accurate calculations. 
To predict values of the slopes with a
precision of several percent one
needs computer solutions of eqs. (1), (2) with the
refinements mentioned at the end of section~\ref{sec-means}.
At present, analytical estimates provide accuracy on
the order of 10--20\%. 
By itself, it is astonishing to obtain any agreement at all
given the large value of the expansion parameter 
$\gamma _0\approx 0.5$ at hand.
This should be considered as a success of the whole approach.

The predictions about curvatures and widths,
as well as higher moments, 
should also be confronted to experiment (see \cite{dr4, and, ede, dln}).

\section{Summary}

We have shown that purely perturbative QCD estimates of the anomalous
dimension $\gamma $ and the ratio of average multiplicities in gluon and
quark jets $r$ are quite reliable at the level of about 10-20$\%$,
where the nonperturbative corrections become important. 
The simplest estimates of these corrections arising from the 
limitation of the parton cascade by some virtuality cut-off are given.

For the slope of $r$ (i.e., $r^{'})$
and for the ratio of the logarithmic slopes of average multiplicities $\rho_1$,
the perturbative results become reliable only at much larger energies than
the $Z^0$. However, the qualitative trends of the 
theoretical corrections demonstrate a tendency to yield
better agreement with experiment as higher orders are included. 
In particular, the partial compensation of higher order corrections 
in the ratio of the slopes of average multiplicities $r^{(1)}$ results 
in a good description of the data by the lowest NLO-expression.
These conclusions are obtained using corrections of order
$\gamma _{0}^{3}$  in $\gamma $ and $r$ as given by the equations for
the generating functions. To evaluate the slopes at the $Z^0$ and below with
better accuracy one should solve the equations numerically.

In the same approximation we have derived the analytical predictions of
perturbative QCD for the curvatures of the average multiplicities of quark and
gluon jets as a function of energy (or of the evolution parameter).
Recent progress in disentangling quark and
gluon jets leads a way to the
experimental verification of these predictions.

\section*{Appendix 1. The anomalous dimension and the ratio of multiplicities}

To obtain the perturbative solution of eqs. (1), (2) 
we use Taylor series expansion
(at large $y$) of multiplicities in the integrals and then differentiate
both sides of the equations. The arguments of $\gamma _0^2$ in the integrals
are chosen to be $y+\ln x(1-x)$ in accordance with the dependence of the coupling
strength on the parton transverse momentum.
Finally, we get the following equations:
\begin{eqnarray}
n_{G}^{''}=\gamma _{0}^{2}\{n_G+2(un_G-n_{G}^{'})\vv_1-(2un_{G}^{'}-n_{G}^{''})\vv_2
+\frac {n_{G}^{'''}}{2}\vv_3+ \nonumber \\
\frac {n_f}{4N_c}[(un_G-n_{G}^{'}-2un_F+2n_{F}^{'})\vv_4
+2(un_{G}^{'}-3un_{F}^{'}+n_{F}^{''})\vv_5+n_{F}^{'''}\vv_6]\},  \label{ang}
\end{eqnarray}
\begin{eqnarray}
r_{0}n_{F}^{''}=\gamma _{0}^{2}[n_G+(un_G-n_{G}^{'})\vv_7+(2un_{G}^{'}-n_{G}^{''})
\vv_8-\frac {1}{2}n_{G}^{'''}\vv_9+  \nonumber   \\
(un_{G}^{'}+un_{F}^{'}-n_{F}^{''})\vv_{10}
-\frac {1}{2}n_{F}^{'''}\vv_{11}],     \label{anf}
\end{eqnarray}
where $u=2B\gamma _{0}^{2}$ and $\vv_i$ are the following integrals with the
limits of integration chosen as $\varepsilon $ and $1-\varepsilon $
evaluated up to terms $O(\varepsilon)$ (see also~\cite{dol}):
\begin{eqnarray}
\vv_1=\int Vdx=\int (1-\frac {3}{2}x+x^2-\frac {x^3}{2})dx=\frac {11}{24}-
\varepsilon, \\
\vv_2=\int \left [\frac {\ln (1-x)}{x}-2V\ln x(1-x)\right ]dx=\frac {67-6\pi ^{2}}
{36}+\varepsilon \ln \varepsilon,  \\
\vv_3=\int \left [\frac {\ln ^{2}(1-x)}{x}-2V(\ln ^{2}x+\ln ^{2}(1-x))\right ]dx=
\nonumber \\
2\zeta (3)-\frac {413}{108}+\varepsilon (\ln ^{2}\varepsilon -2\ln \varepsilon
+2),   \\
\vv_4=\int [x^2+(1-x)^2]dx=\frac {2}{3}-2\varepsilon , \\
\vv_5=\int [x^2+(1-x)^2]\ln x dx=-\frac {13}{18}+\varepsilon (1-\ln \varepsilon ), \\
\vv_6=\int [x^2+(1-x)^2]\ln ^{2}x dx=\frac {89}{54}-\varepsilon (\ln ^{2}\varepsilon
-2\ln \varepsilon +2),  \\
\vv_7=\int \Phi dx=\int (1-\frac {x}{2})dx=\frac {3}{4}-\frac {3}{2}\varepsilon ,  \\
\vv_8=\int \Phi \ln x dx=-\frac {7}{8}+\varepsilon (1-\ln \varepsilon ),  \\
\vv_9=\int \Phi \ln ^{2}x dx= \frac {15}{8}-\varepsilon (\ln ^{2}\varepsilon
-2\ln \varepsilon +2),  \\
\vv_{10}=\int \left [\Phi -\frac {1}{x}\right ]\ln (1-x) dx=\frac {\pi ^2}{6}-
\frac {5}{8}+\frac {1}{2}\varepsilon (\ln \varepsilon -3),  \\
\vv_{11}=\int \left [\Phi-\frac {1}{x}\right ]\ln ^{2}(1-x) dx =\frac {9}{8}
-2\zeta (3)+\frac {1}{2}\varepsilon (\ln ^{2}\varepsilon -2\ln \varepsilon +2),
\end{eqnarray}
$\zeta $ means the $\zeta $-function, with $\zeta(3)=1.2021$.

Using formulas (\ref{gam}), (\ref{rat}) and keeping terms up to $O(\gamma _{0}
^{5})$ order on both sides of eqs. (\ref{ang}), (\ref{anf}) we get
\begin{eqnarray}
a_1=\vv_1-\frac {B}{2}+\frac {n_f}{8N_c}\left (1-\frac {2}{r_0}\right )\vv_4,  \\
a_2=\left (\frac {a_1}{2}-\vv_1-\frac {n_f}{12N_c}(1-\frac {2}{r_0})\right )
(a_1+2B)-\frac {\vv_2}{2}-\frac {n_f}{12N_{c}r_0}(2r_1+3\vv_5),  \\
a_3=a_2(a_1-\vv_1+\frac {3}{2}B)-\frac {1}{2}BB_1+(a_1+\frac {5}{2}B)\vv_2-
\frac {1}{4}\vv_3-\frac {n_f}{8N_c}\{ [a_2\left (1-\frac {2}{r_0}\right )-
\nonumber   \\
\frac {2}{r_0}(a_1r_1+3Br_1-r_{1}^{2}-r_2)]\vv_4+2\left [2B-(2a_1+7B-r_1)\frac
{1}{r_0}\right ]\vv_5+\frac {1}{r_0}\vv_6\},
\\
r_1=2\vv_1-\vv_7+\frac {n_f}{4N_c}\left (1-\frac {2}{r_0}\right )\vv_4,
\\
r_2=r_1(\vv_7-a_1-\frac{n_f}{3N_{c}r_0})-\vv_2-\vv_8 - \frac {1}{r_0}\vv_{10}
\nonumber \\
+ \frac{n_f}{12N_{c}r_0}
[(r_0-2)(a_1+2B)(3\vv_4-2)
-6\vv_5],  \label{r2v}\\
r_3=2a_3-2(a_2-r_2)(a_1-r_1)-r_1[(a_1-r_1)^2+4Ba_1+3B^2]+
B[B_1+2\vv_{10}\nonumber   \\-3a_2+5r_2]+
a_2\vv_7+(2a_1+5B)\vv_8-\frac {1}{2}\vv_9+\frac {1}{r_0}[(2a_1+3B-r_1)\vv_{10}
-\frac {1}{2}\vv_{11}] + 5Br_1^2.
\end{eqnarray}
The numerical values of $a_i, r_i$ are given in Table~\ref{tab-randa} for
$\varepsilon =0$. 
The corresponding curves for $r$ using $n_f=4$ are shown in Fig. 1.

The formulas for $a_1, a_2, r_1$ coincide with those in \cite{mu, mw}.
The expression for $r_2$ was obtained in \cite {dr3}. 
Use of Taylor series expansion allows energy conservation to be included in
the vertices of Feynman graphs. This important fact reveals itself in a special
form of the integrals $\vv_2, \vv_{10}$ containing $\pi ^2$. Just such a form of
corrections was also obtained in the Lund model \cite{ede} when taking into
account energy conservation by cutting off the fractal triangles in the
($y, k_T$)-variables.

Earlier, in the renormalization group approach~\cite{gaff},
only part of our expression for $r_2$ was obtained
(the terms with $r_1$ in (\ref{r2v})),
which is 6 times smaller than the other terms. 
It is difficult to account for energy conservation in that approach.

The rather large value of $r_2$ compared with $r_1$ raised suspicion 
about the nature of the series for $r$ and its convergence,
even at the $Z^0$. 
However, the much smaller value of $r_3$ suggests that 
this is just a numerical fluctuation, and that the series is well 
behaved and can be trusted. 
This is supported by results of the computer calculations~\cite{lo}.

\section*{Appendix 2. The slopes}

The slope of the ratio of multiplicities is given by eq. (\ref{rpri}).
Using the constants in Table~\ref{tab-randa} for $n_f=4$ we get
\begin{equation}
r^{'}\approx 0.16\gamma _{0}^{3}(1+4.9\gamma _{0}+1.5\gamma _{0}^{2}+O(
\gamma _{0}^{3})).   \label{rnum}
\end{equation}
The factor in front of the bracket is very small in the available energy
region with $\gamma _{0}\sim 0.5$. The second term is, however,
quite large compared to 1 at present energies, even though it
vanishes asymptotically. Thus one can hardly
trust the perturbative expression for $r^{'}$ at the $Z^0$
and can use it for qualitative estimates only.

The perturbative expression for $\rho _1$ is
\begin{eqnarray}
\rho_1 =
1-Br_1\gamma _{0}^{2}[1+(a_1+r_1+\frac{2r_2}{r_1})
\gamma _0+ \nonumber \\
(\frac {2r_2a_1}{r_1}+a_1r_1+3r_2+\frac{3r_3}{r_1}+a_2+a_{1}^{2}
+r_{1}^{2}+B_1)\gamma _{0}^{2}].   \label{d-1}
\end{eqnarray}
The correction to 1 is small and vanishes rapidly as the scale increases.
However, the subsequent terms in the brackets are not small
at present energies as seen from their numerical values
presented in eq. (\ref{cor3}) for $n_f=4$.

The situation is similar for the ratio of the curvatures (second derivatives)
of multiplicities, which has the solution
\begin{eqnarray}
\rho_2 =
1- 2B r_1\gamma^2_0 [1 + (\frac{2r_2}{r_1}+r_1+a_1-\frac{B}{2})\gamma_0
\nonumber \\
+(\frac{3r_3}{r_1}+3r_2+r_1^2+a_2+\frac{2a_1r_2}{r_1}+a_1r_1+a_1^2+B_1
\nonumber \\
-\frac{2Br_2}{r_1}-\frac{3}{2}Br_1-2Ba_1-\frac{B^2}{2})\gamma_0^2
]. \label{d-2}
\end{eqnarray}
Numerically (for $n_f=4$),
\begin{equation}
\rho _2 =
1-0.14\gamma^2_{0}(1+5.21\gamma _{0}+2.14\gamma _{0}^{2}).    \label{nd2}
\end{equation}
Once again the terms in the brackets are larger or compatible with 1
at present values of $\gamma _0 \sim 0.5$. 
All the above expressions can be trusted quantitatively
only at extremely high
energies where $\gamma _0\leq 0.2$, i.e. $\alpha _S\leq 0.02$,
far below the present value $\alpha _S\approx 0.12$.

Thus we conclude that purely perturbative QCD as described by equations
(1), (2) for the multiplicities with integration limits equal to 0 and 1 can
hardly be accepted for quantitative estimates of the logarithmic slopes.
Qualitatively, it predicts a positive slope for $r$ and an excess of the ratio
of slopes $r^{(1)}$ over the ratio $r$,
in accordance with experimental findings. However their numerical
values are lower than the experimental ones if only terms quoted above are
taken into account.

\section*{Acknowledgements}

This work has been supported by INTAS (V.N. -- grant 97-103),
by the Russian Fund for Basic Research,
and by the US Department of Energy under grant DE-FG03-94ER40837.
One of us (A.C.)
would like to thank Yu.~Dokshitzer, M.~Fontannaz, A.~Kaidalov,
G.~Korchemski, A.~Krzywicki and D.~Schiff for useful discussions.

\newpage
\begin{figure}[p]
\begin{center}
  \epsfxsize=14cm
  \epsffile[25 0 555 600]{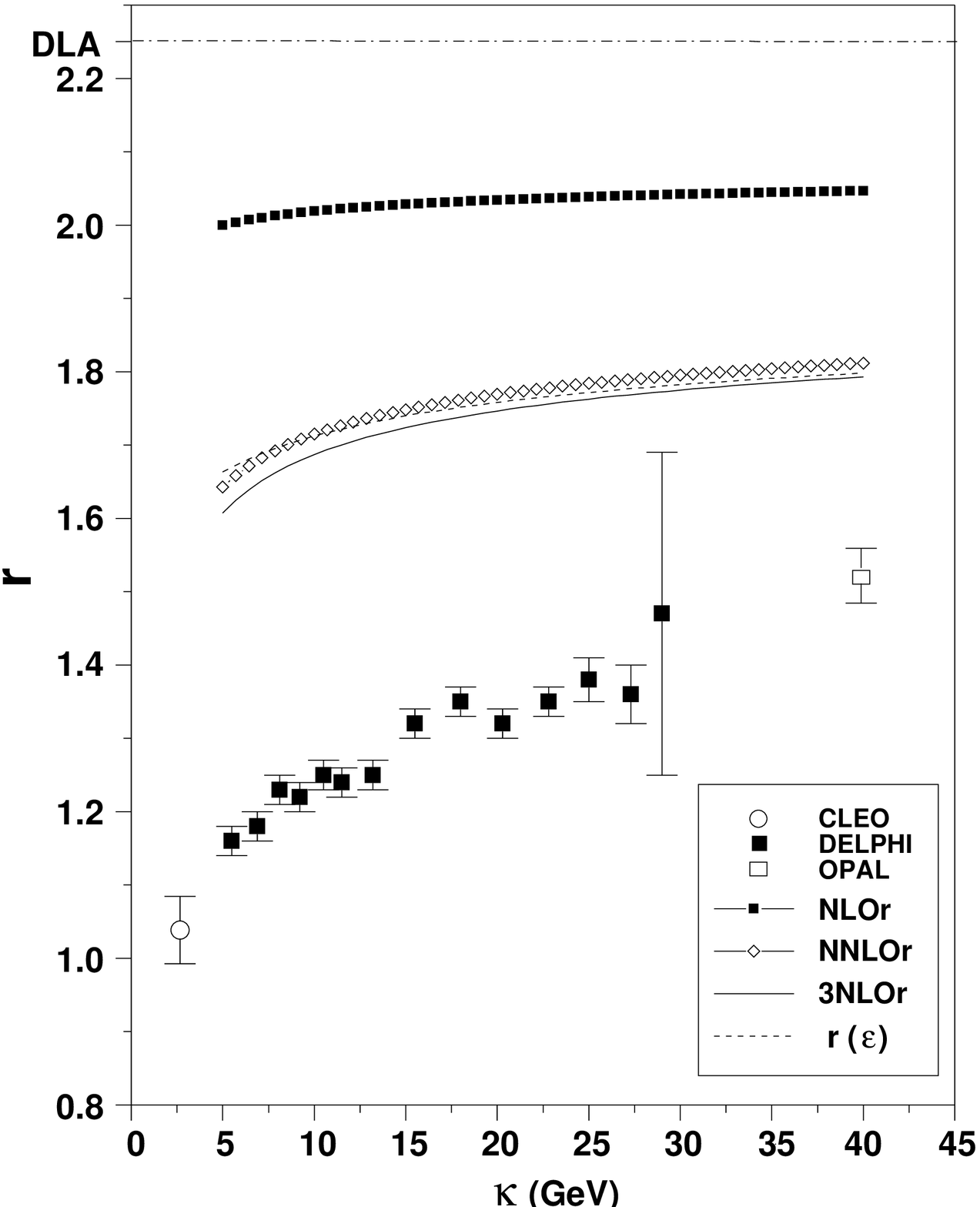}
\end{center}
\vspace*{-.5cm}
\caption{
The ratio of the average multiplicities in gluon and
quark jets versus the scale $\kappa$ in the DLA, NLOr,
NNLOr and 3NLOr approximations using integration limits 0--1, 
and in the 3NLOr approximation
using limits $e^{-y}$ and $1-e^{-y}$ 
($r(\varepsilon )$ line),
in comparison to data~\cite{cleo97,opal99,lan}.
The theoretical results are obtained using $n_f=4$.
} 
\end{figure}

\newpage
\begin{figure}[p]
\begin{center}
  \epsfxsize=14cm
  \epsffile[25 0 555 600]{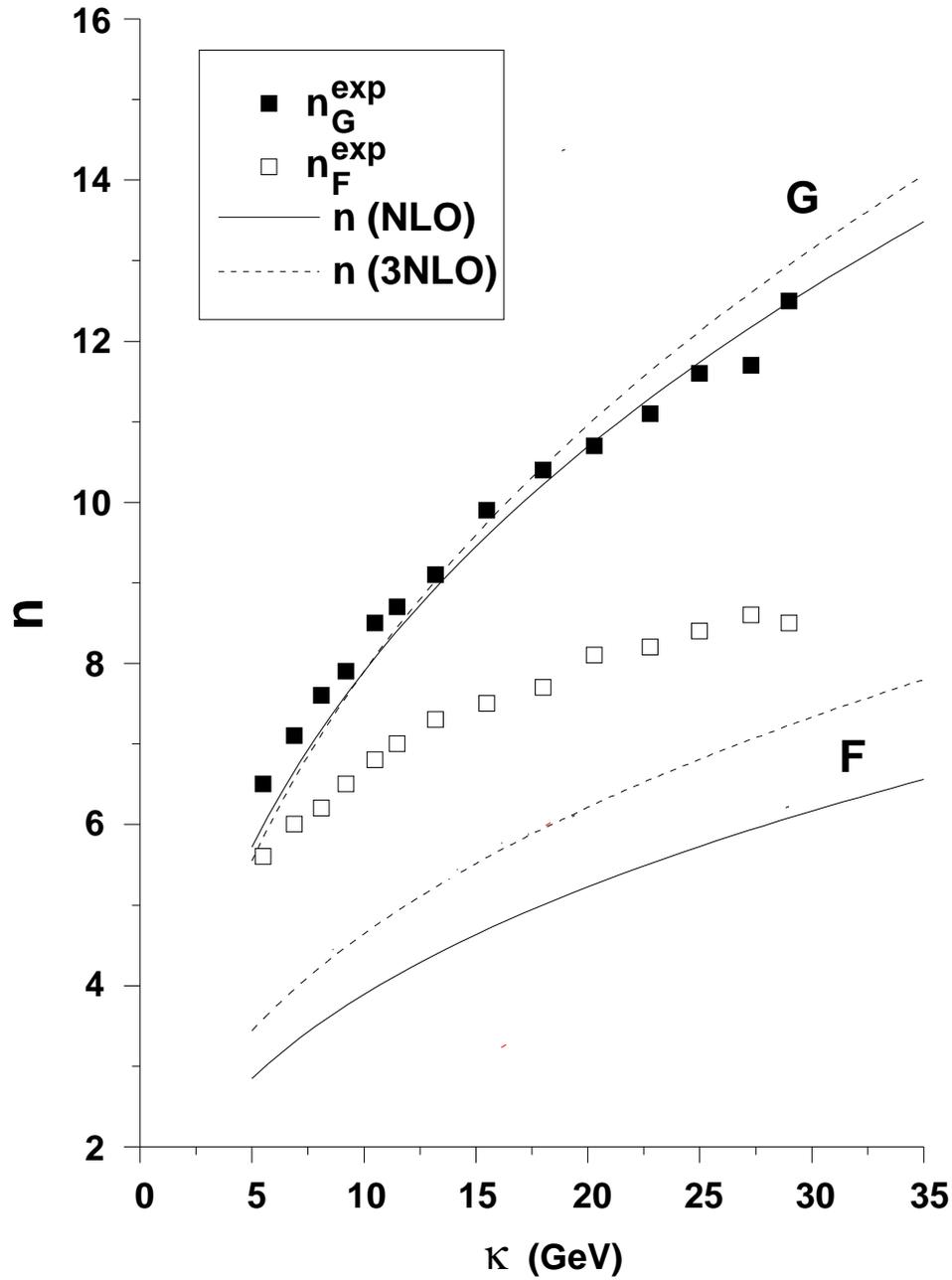}
\end{center}
\vspace*{-.5cm}
\caption{
The average multiplicities of gluon (G) and quark (F) jets
in the NLO and 3NLO approximations,
compared to data~\cite{lan}.
For the theory,
the gluon jet normalization is fit to the data;
the normalization of the gluon jet curve 
fixes the normalization of the quark jet curve.
The theoretical results are obtained using $n_f=4$.
} 
\end{figure}

\newpage
\begin{figure}[p]
\begin{center}
  \epsfxsize=14cm
  \epsffile[25 0 555 600]{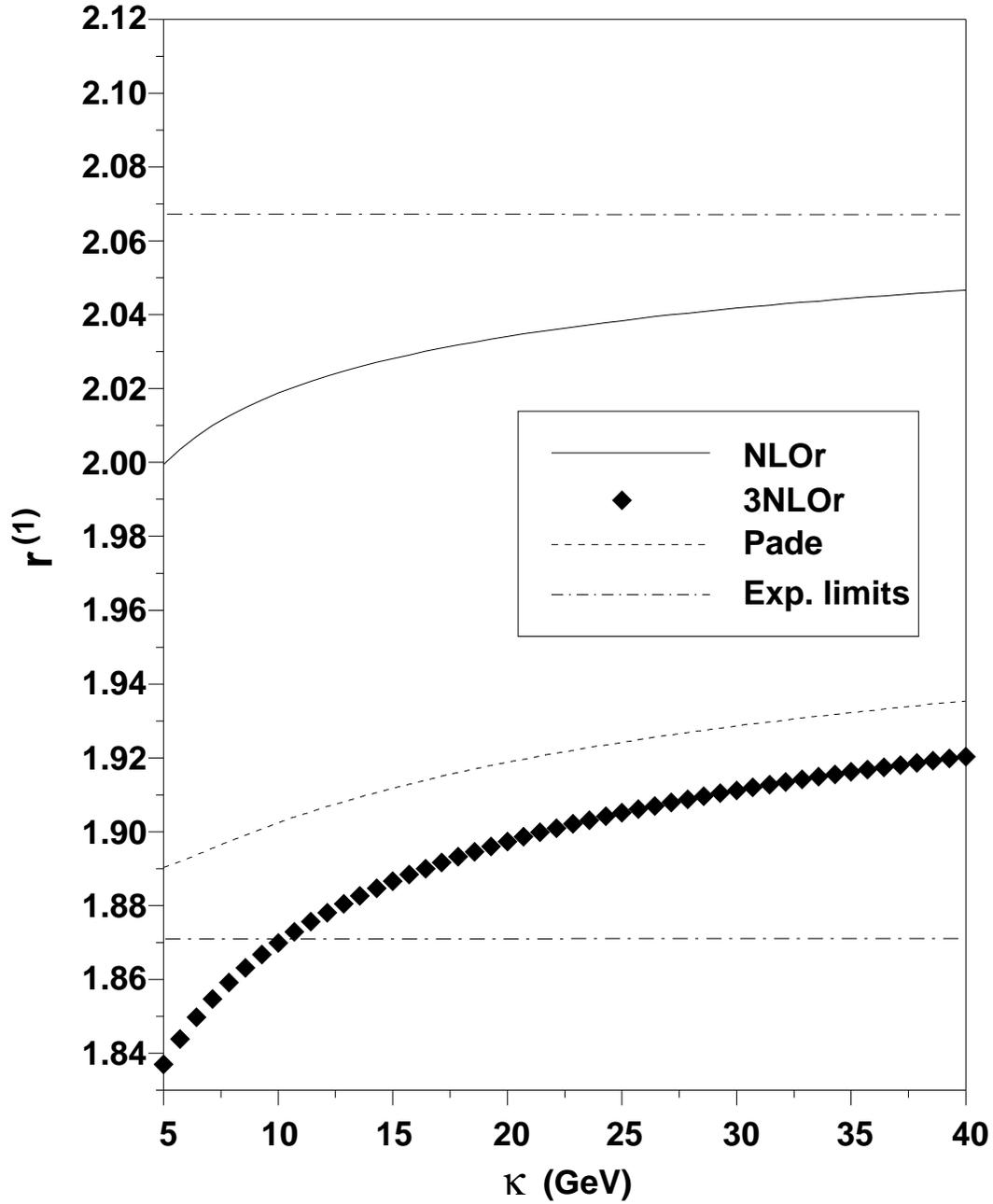}
\end{center}
\vspace*{-.5cm}
\caption{
The ratio of slopes of the average multiplicities 
in gluon and quark jets, $r^{(1)}$,
in the NLOr and 3NLOr (with its Pade expression) approximations.
The theoretical results are obtained using $n_f=4$.
Upper and lower bounds of experimental results~\cite{lan} 
are shown by the dash-dotted lines.
} 
\end{figure}

\end{document}